\documentstyle[eqsecnum,prd,aps,a4]{revtex}
\input epsf

\def\prl{Phys.\ Rev.\ Lett.}
\def\pr{Phys.\ Rev.}

\def\np{Nucl.\ Phys.}

\def\camup{Cambridge University Press}

\begin{document}
\draft
\title{Evolution of Superconducting String Currents}
\author{C. J. A. P. Martins\thanks{Also at C. A. U. P.,
Rua do Campo Alegre 823, 4150 Porto, Portugal.
Electronic address: C.J.A.P.Martins\,@\,damtp.cam.ac.uk}
and
E. P. S. Shellard\thanks{Electronic address:
E.P.S.Shellard\,@\,damtp.cam.ac.uk\vskip0pt Submitted to
Phys.\ Lett.\ {\bf B}.}}
\address{Department of Applied Mathematics and
Theoretical Physics\\
University of Cambridge\\
Silver Street, Cambridge CB3 9EW, U.K.}
\date{22 May 1997}
\maketitle

\begin{abstract}
We extend the quantititative string evolution model of Martins and Shellard
to superconducting strings by introducing a simple toy model for the
evolution of the currents. This is based on the dynamics of a `superconducting
correlation length'. We derive the relevant evolution equations and discuss
the importance of plasma effects. Some consequences of our results are
also suggested.
\end{abstract}

\section{Introduction}
\label{v-in}
It is well known that cosmic strings can in
some circumstances (typically when the electromagnetic gauge invariance is
broken inside the string) behave as
`superconducting wires' carrying large currents and charges---up to the
order of the string mass scale in appropriate units \cite{witten}.
The charge carriers can be either bosons or fermions (see \cite{vs} for a
review). The former type occurs when it becomes energetically
favourable for a charged Higgs field to have a
non-zero vacuum expectation value in the string core; the latter happens
when there are fermion zero modes moving in the string field.

If superconducting strings carry currents, they must also carry charges of
similar magnitude. This includes not only charges trapped at formation by
the Kibble mechanism but also the ones due to string inter-commuting
between regions of the string network with different currents.
However, arbitrarily large currents or charges are not allowed---there are
critical values beyond which the current saturates and charge carriers
can leave the string.

Even though the overwhelming majority of the work done on cosmic strings so far
was concerned with the structureless Goto-Nambu strings, there are a number
of cosmological scenarios in which superconducting strings play a crucial
part---notably the so-called `vorton problem' \cite{vor}---so that
their study is highly relevant.

In this paper we take a first step in this direction by generalizing
the quantitative string evolution model of Martins and
Shellard \cite{ms,ms1,ms2,ms2a,mac}. After a brief review of this model,
we extend it by developing a simple toy model for the evolution of the
currents on the long strings, based on the dynamics of a `superconducting
correlation length'. The different contributing
mechanisms are discussed, and in particular we consider the importance
of plasma
effects. Finally, we briefly mention some of the implications
of our model---a more detailed discussion is left for a forthcoming
publicatiom \cite{msmag}.
 
Throughout this paper we will use fundamental units in which
$\hbar=c=k_B=Gm^2_{Pl}=1$.

\section{Evolution of the Currents}
\label{v-cr}
Analytic string evolution methods describe the string network by a small number
of macroscopic quantities whose evolution equations are derived from the
microscopic string equations of motion. The first such
model providing a quantitative picture of the
complete evolution of a string network (and the corresponding loop
population) has been recently developed by Martins and
Shellard \cite{ms,ms1}, and has two such quantities, the long-string
correlation length $\rho_{\infty}\equiv\mu/L^2$ ($\mu$ being the string
mass per unit length) and the string RMS velocity,
$v^2\equiv\langle{\dot{\bf x}}^2\rangle$.
It also includes two `phenomenological' terms, a `loop chopping
efficiency' parameter $0<{\tilde c}<1/2$ and a `small-scale structure
parameter' $0<k<1$. Their evolution equations are
\begin{equation}
2 \frac{dL}{dt}=2HL(1+v_\infty^2)+
v_\infty^2\frac{L}{\ell_f}+{\tilde c}v_\infty \, , \label{evl}
\end{equation}
\begin{equation}
\frac{dv}{dt}=\left(1-v^2\right)\left[\frac{k}{L}-v\left(2H+
\frac{1}{\ell_f}\right)\right] \, ;
\label{evv}
\end{equation}
these are sufficient to quantitatively
describe the large-scale properties of a cosmic string network.

Note that the explicit form of the `friction lengthscale' $\ell_{\rm f}$ will
depend on the type of current involved. For the case of a neutral current, one
expects Aharonov-Bohm scattering \cite{rsa} to be the dominant effect,
and consequently we have \cite{ms1}
\begin{equation}
\ell_{\rm f}=\frac{\mu}{\beta T^3}\, ,
\label{frilen}
\end{equation}
where $T$ is the background temperature and $\beta$ is a numerical factor
related to the number of particle species interacting with the string.
We assume that this is the case for the time being, but the more general
case where plasma effects can be important will be discussed towards the end
of this paper.

We now extend it setting up a `toy model' for the evolution
of the currents. One assumes that there is a `superconducting
correlation length', denoted $\xi$, which measures the scale over which
one has coherent current and charge densities on the strings. Associated with
this we can define $N$ to be the number of uncorrelated current regions
(in the long-string network) in a co-moving volume $V$, and it is then fairly
straightforward to see how the dynamics of the string network affects $N$ and
obtain an evolution equation for it.

\begin{figure}
\vbox{\centerline{
\epsfxsize=0.6\hsize\epsfbox{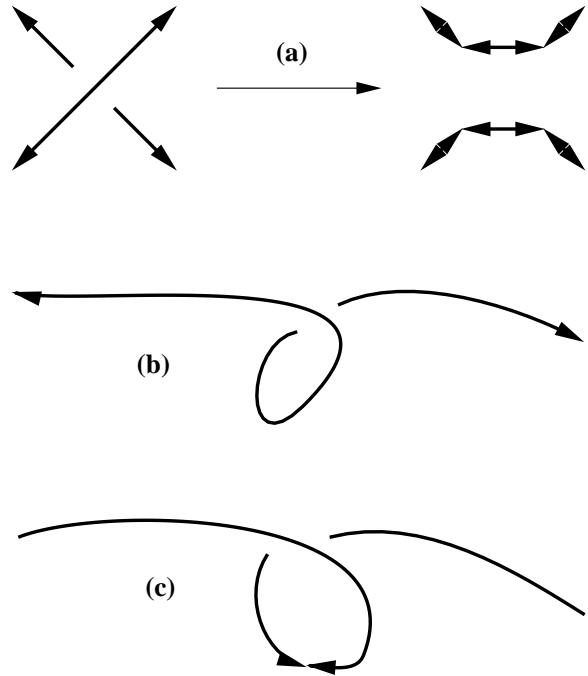}}
\vskip.4in}
\caption{Some relevant inter-commuting configurations. The arrows mark the
limits of regions with correlated currents. Plot (a) shows a typical
inter-commuting creating four new current regions, while (b-c) show than
on scales smaller than the current correlation length loop production may (c)
or may not (b) remove current regions from the long-string network.}
\label{bonec}
\end{figure}

Firstly, we expect that in a co-moving volume the number of uncorrelated
regions will not be affected by expansion.
Now consider the effect of inter-commutings (whether or not a loop is produced).
Their effect on $N$ can easily be obtained by multiplying the inter-commuting
rate (obtained from the original analytical model \cite{ms1}) by the number
of regions created by each inter-commuting---this will obviously be a
positive term.
Laguna and Matzner \cite{laguna} have numerically shown that whenever two
current-carrying strings cross, they inter-commute and a region of
intermediate current is created. This means that inter-commutings
will in general create four new regions (see figure \ref{bonec} (a)). An
exception to this is that when regions with size of order $\xi$ or smaller
self-intersect it is possible (see figure \ref{bonec} (b-c)) that no new
regions are produced. Also, when the inter-commuting does produce a loop, the
regions in the corresponding segment are removed from the network, together
with one of the newly created `intermediate' regions, so that there will
be a negative term in the evolution of $N$ (with a similar correction factor
at small scales).

Finally, the only non-trivial issue is that of
the dynamics of the currents themselves. The simulations of
Laguna and Matzner \cite{laguna} show that as the result of inter-commutings
charges pile up at current
discontinuities and move with the kinks, but their strength decreases with
time. Clearly, this indicates that some kind of `equilibration' process
should act between neighbouring current regions, which will counteract the
creation of new regions by inter-commuting (and help the removal of regions
by loops). Furthermore, Austin, Copeland
and Kibble have shown \cite{ack} that in an expanding universe correlations
between left- and right-moving modes develop due both to stretching and
inter-commuting (particularly when loops form). We model this term by
assuming that after each Hubble time, a fraction $f$ of the $N$ regions
existing at its start will have equilibrated with one of its neighbours,
\begin{equation}
\left(\frac{dN}{dt}\right)_{dynamics}=-fHN\, ;
\label{eqlterm}
\end{equation}
note that new regions are obviously created by inter-commuting during the
Hubble time in question, so that $f$ can be larger than unity.
Alternatively we can say that for a given $f$, the number of regions that were
present in a given volume at a time $t$ will have disappeared due to
equilibration at a time $t+(fH)^{-1}$. Presumably the only way to find out
what $f$ is is by means on numerical simulations (in particular, we would
expect it to be a model-dependent quantity), although some physical arguments
can be used to constrain it \cite{msmag}.

However, setting this issue aside for the time being, we
obtain the following evolution equation for $N$
\begin{equation}
\frac{dN}{dt}=G\left(\frac{\ell}{\xi}\right)
\frac{v_\infty}{\alpha}\frac{V}{L^4}-fHN\, ,
\label{generalnn}
\end{equation}
where the `correction factor' $G$ has the form
\begin{equation}
G\left(\frac{\ell}{\xi}\right)=\left\{ \begin{array}{ll}
2-{\tilde c}\left(\frac{\ell}{\xi}+2\right) \, ,&
\mbox{$\frac{\ell}{\xi}>1$} \\
2(1-2{\tilde c})\alpha+(2-3{\tilde c}-2\alpha+4\alpha{\tilde c})
\frac{\ell}{\xi} \, ,& \mbox{$\frac{\ell}{\xi}\le1$} \end{array} \right. \, ;
\label{gansg}
\end{equation}
loops are assumed to form with a size $\ell(t)=\alpha(t)L(t)$, with
$\alpha\sim1$ while the string network is is the friction-dominated epoch
and $\alpha=\alpha_{sc}\ll1$ once it has reached the linear scaling regime
(see \cite{ms1}). Note that when $\ell\gg\xi$ the net effect of
inter-commuting and loop production is to remove uncorrelated regions
(because each loop formed removes a large number of them); otherwise,
the net effect is to create new regions.

For what follows it is more convenient to introduce $N_L$, defined to be
the number of uncorrelated current regions per long-string correlation length,
\begin{equation}
N_L\equiv\frac{L}{\xi}\, ;\label{defnl}
\end{equation}
in terms of $N_L$, (\ref{generalnn}) has the form
\begin{equation}
\frac{dN_L}{dt}=(3v_\infty^2-f)HN_L+\frac{3}{2}
\frac{v_\infty^2}{\ell_{\rm f}}N_L+\left(\frac{1}{\alpha}G(\alpha N_L)+
\frac{3}{2}{\tilde c}N_L\right)\frac{v_\infty}{L}\, ,
\label{generalnl}
\end{equation}
(with the obvious definition for $G(\alpha N_L)$) where $\ell_{\rm f}$ is the
relavant friction lengthscale. Note that to obtain this
equation one needs to use the evolution equation for the long-string
correlation length $L$.

We should also say at this stage that once the network leaves the
friction-dominated regime and strings become relativistic other mechanisms
(notably radiation) can cause charge losses. Thus we do not expect our toy
model to provide quantitatively correct answers in this regime, but we do
expect it to provide reliable order-of-magnitude estimates.

\section{Equilibration and Scaling}
\label{v-ff}
We start by discussing the case of neutral currents, in which $\ell_{\rm f}$
is given by (\ref{frilen}).
Analysis of (\ref{generalnl}) together with the evolution equations for
the long-string network reveals two types of behaviour.
Firstly, if $f$ is small (that is, equilibration is ineffective) then $N_L$
grows without limit. In the particular case $f=0$, $N_L\propto t$ once
the long-string network has reached the linear scaling regime, meaning that
$\xi\propto const.$
On the other hand, for
\begin{equation}
f>f_{min}=3-\frac{2{\tilde c}}{k+{\tilde c}}\sim1.88\, \label{fmins}
\end{equation}
(where we have used the scaling values of ${\tilde c}$ and $k$ obtained
from numerical simulations in the radiation era---see \cite{ms1} for
a discussion)
the late time behaviour is $N_L\propto const.$, $\xi\propto t$, that is
the superconducting correlation length is scaling linearly just like the
long-string correlation length; the scaling value of $N_L$ has a fairly
weak dependence on $f$, with smaller $N_L$'s corresponding to larger $f$'s
as expected.

Note that the fact that correlations cannot obviously be established
faster than the speed of light (that is, we must have $\xi\le t$), means
that there is a maximum value allowed for $f$, which at late times (in the
radiation era) can be written
\begin{equation}
f<f_{max}=3+\frac{4(1-2{\tilde c})}{k^{1/2}(k+{\tilde c})^{3/2}}\sim22.4
\, .\label{fmaxs}
\end{equation}

\section{Two Examples}
\label{v-rs}
\begin{figure}
\vbox{\centerline{
\epsfxsize=0.6\hsize\epsfbox{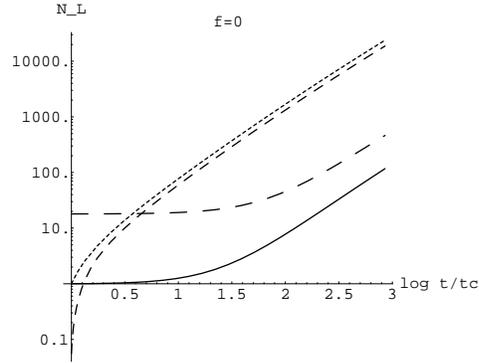}}
\vskip.4in}
\vbox{\centerline{
\epsfxsize=0.6\hsize\epsfbox{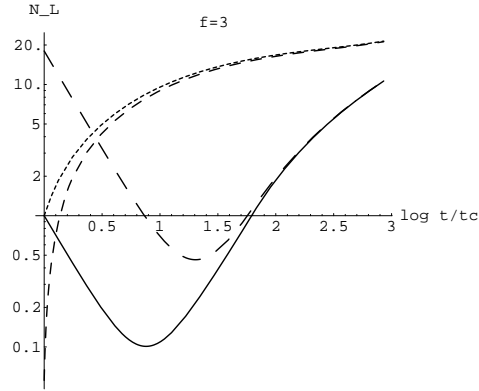}}
\vskip.4in}
\caption{The evolution of the number of uncorrelated current regions per
long-string correlation length, $N_L$, for the cases $f=0$ (top) and
$f=3$ (bottom) assuming that
the orders of the string-forming
and superconducting phase transitions are respectively:
1st \protect\& 1st (solid lines), 1st \protect\& 2nd (long dashed),
2nd \protect\& 1st (short dashed) and 2nd \protect\& 2nd (dotted). Time is in
orders of magnitude from the epoch of string formation.}
\label{evn_l3}
\end{figure}

In figure \ref{evn_l3} we plot the result of the numerical
integration of (\ref{generalnl}) for GUT string networks in the
friction-dominated epoch, for initial conditions
representative of string-forming and superconducting phase transitions of
first and second order, for the cases $f=0$ and $f=3$, which are
representative of the cases where equilibration is ineffective and effective,
respectively. We are assuming that
both of these phase transitions occur at around the same
(GUT) energy scale. It was also assumed that
the value of $\alpha$ in the linear scaling regime is $\alpha_{sc}\sim10^{-3}$
(see Martins and Shellard \cite{ms1}).

The differences between the two cases are considerable. If there is no
equilibration mechanism ($f=0$), $\xi$ is conformally stretched during the
stretching regime (just like the long-string correlation length, $L$), and
so $N_L$ is approximately constant. However, as inter-commutings start
creating new regions $N_L$ begins to increase, growing as $t^{3/2}$ during
the Kibble regime and eventually (once the network
reaches the linear scaling regime) ends up growing as $N_L\propto t$, which
corresponds to $\xi\propto const$. As expected, in this case the network
keeps a `memory' of its initial conditions.
On the other hand, if there is an
equilibration mechanism ($f=3$) then $N_L$
decreases while the network is being conformally stretched. In the Kibble
regime, the increased number of inter-commutings again drives $N_L$ up,
and after $\alpha$ has evolved into its linear regime value $\xi$ itself
reaches a scaling value and hence $N_L$ becomes a constant.

\section{Plasma Effects}
\label{v-pl}
If the cosmic strings interact with a plasma there will be a further damping
term, with the corresponding friction lengthscale being given by
\begin{equation}
\ell_{\rm j}=\frac{\mu}{\rho^{1/2}J}\, ,
\label{frip}
\end{equation}
where $\rho$ is the plasma density and $J$ is the string current; at a scale
$r$ this is given by
\begin{equation}
J(r)=e\frac{N_r^{1/2}}{r}\, ,
\label{jrr}
\end{equation}
$N_r$ being the number of uncorrelated current regions at that scale.

We must therefore determine which of the two damping lengthscales is dominant
(that is, smaller). It is straightforward to find that at a time $t=xt_c$
($t_c$ being the epoch of string formation)
\begin{equation}
\frac{\ell_{\rm j}}{\ell_{\rm f}}=\left(\frac{32\pi}{3}\right)^{1/2}
\frac{\gamma x^{1/2}}{deN_L^{1/2}(G\mu)^{1/2}}\, ,
\label{frirad}
\end{equation}
where $L=\gamma t$ and $d=4\pi(\pi {\cal N}/45)^{1/2}$, (${\cal N}$ being the
number of effectively massless degrees of freedom). We therefore require this
ratio to be less than unity for plasma effects to be dominant.

Now, what initial currents do we expect at $t_c$? If plasma effects are to be
important, $\ell_{\rm j}<\xi<t$ at $t_c$; note that $\ell_{\rm j}$ depends
on $\xi$, but we can still solve consistently for the minimum allowed
$\ell_{\rm j}$, which will produce the largest current. Using these bounds we
find in the GUT case
\begin{equation}
4\le\left(\frac{\ell_{\rm j}}{\ell_{\rm f}}\right)_{tc}\le60\, ,
\label{frilims}
\end{equation}
so that plasma effects are initially subdominat;
the lower bound is $G\mu$-independent, whereas the upper one varies as
$(G\mu)^{-1/2}$. However, note that the lower bound depends on ${\cal N}$, 
with a ratio of four corresponding to a minimal GUT model ${\cal N}\sim106.75$;
one would need ${\cal N}\sim10^4$ for plasma effects to become dominant---but
even so, this would only happen in the `extreme' case of a strongly first
order string-forming phase transition and a second order superconducting
phase transition.

Thus we can in general study the initial stages of superconducting string
evolution with (\ref{generalnl}) and the Aharonov-Bohm friction lenghtscale.
But then the evolution of the currents is such that the ratio (\ref{frirad})
is at most approximately constant---this happens in the `extreme' case
of no equilibration, $f=0$, in which $N_L$ respectively behaves as
$N_L\propto const.$, $N_L\propto t^{3/2}$ and $N_L\propto t$ in the
stretching, Kibble and linear scaling regimes. If there is equilibration
(that is $f\neq0$) then $N_L$ grows more slowly than the above and the
plasma damping effect loses importance relative to Aharonov-Bohm scattering
(that is, $\ell_{\rm j}$ grows faster than $\ell_{\rm f}$).

Hence we find that plasma effects are in general subdominant. However, a way
of avoiding these constraints is to have further current-building mechanisms,
such as background magnetic field---this is in fact commonly assumed (and not
very well explained) in most of the existing litterature \cite{chud}. What
we have shown is that if no such charging mechanism exists than the currents
on the strings are much smaller.

\section{Conclusions}
\label{v-cc}
In this paper we have constructed a simple `toy model' for the evolution
of currents on cosmic strings, based on the dynamics of a `superconducting
correlation length. The different dynamical processes affecting it were
considered, and a simple study of the solutions of the model has
revealed that the process of equilibration is crucial in the evolution of
the currents. We have found that a fairly efficient equilibration mechanism
is needed for the superconducting correlation length to scale. Whether or
not such a mechanism is available in realistic model is an issue which cannot
be addressed at the moment---a detailed numerical investigation will
clearly be needed for that.

We have also studied the importance of plasma damping effects on the strings.
We find that these are typically negligible when compared to the `usual'
Aharonov-Bohm scattering. However, an exception to this might happen in the
presence of background magnetic fields, either of `primordial' origin or
generated (by a dynamo mechanism, say) after protogalaxies start to collapse
(an example of the later is the Chudnivsky-Vilenkin scenario \cite{chud}).
Even in this case, for typical values of cosmological or astrophysical
magnetic fields, the two damping lengthscales are not too different.

While this paper was being prepared, a rather less accurate discussion of
plasma effects on string dynamics appeared in the litterature \cite{dd}.
This is in broad agreement with our results if one allows for the fact that
in \cite{dd} no consideration is given to the dynamics of currents---in fact
current is simply treated as a constant.

Our extended quantitative model allows reliable estimates of the currents
on strings to be made at all times. Thus a more detailed analysis of some
of the cosmological scenarios involving superconducting cosmic strings
becomes possible. An outstanding example the question of the abundance
of vorton remnants---we will address it in a forthoming publication\cite{msmag}.

\vfill\eject

\end{document}